\documentclass[aps,prb, twocolumn, superscriptaddress, longbibliography, nobalancelastpage]{revtex4-1}

\usepackage{times}
\usepackage{amsmath}
\usepackage{amssymb}
\usepackage{graphicx}
\usepackage[caption=false, position=top, singlelinecheck=off, justification=raggedright]{subfig}
\usepackage{color}
\usepackage{pst-node}
\usepackage[unicode]{hyperref}
\definecolor{ACSblue}{RGB}{13,84,166}
\hypersetup{unicode=true, colorlinks=true, linkcolor=ACSblue, citecolor=ACSblue,urlcolor=ACSblue }
\usepackage{float}
\usepackage[normalem]{ulem}
\usepackage{multirow}
\usepackage{hhline}

\usepackage{mathrsfs,braket}
\usepackage{amssymb, amsbsy, amsmath, latexsym, dsfont, array, layout, graphicx, mathrsfs, color, ulem, bm}

\usepackage{changepage}  

\usepackage{xcolor}
\usepackage{tikz}
\definecolor{ACSblue}{RGB}{38,51,128}

\newcommand{\ACsquare}{
  \tikz[baseline=0ex]{
    \fill[ACSblue] (0,0) rectangle (0.8em,0.8em);
  }
}
\newcommand{\SIicon}{
  \tikz[baseline=(SI.base)]{
    \node[draw,circle,fill=black,inner sep=1pt] (SI)
      {\sffamily\bfseries\color{white}\scriptsize SI};
  }
}

\usepackage[T1]{fontenc}
\usepackage[scaled=1.0]{sourcesanspro}

\newcommand{\ACSFont}[1]{
  {
    \fontfamily{SourceSansPro-TLF}\fontseries{sb}\selectfont 
    \large 
    \textcolor[RGB]{38,51,128}{#1}%
  }
}

\newcommand{\ACSection}[1]{
  \noindent
  \hspace*{-0.4em} \ACsquare\hspace{0.01em} 
  \ACSFont{#1}\par
  \vspace{0.3\baselineskip}
}

\newcommand{\ACSSISubtitle}[1]{
  {
\fontfamily{SourceSansPro-TLF}\fontseries{sb}\selectfont
   \normalsize #1}
}
 

\makeatletter
\renewcommand*{\fnum@figure}{\textbf{Figure \thefigure}}

\renewcommand*{\@caption@fignum@sep}{\textbf{.} }
\makeatother

\definecolor{orcidlogocol}{HTML}{A6CE39}

\newcommand{\orcidicon}{
  \tikz[baseline=-0.6ex]{
    \node[draw=orcidlogocol, fill=orcidlogocol,
          circle, inner sep=0pt, minimum size=1.6ex]
         {\scriptsize\textcolor{white}{iD}};
  }
}

\begin{document}

\title{ Synergy and Competition of Dual Chirality in the Chirality-Induced Spin Selectivity of Supramolecular Helices}

\author{Song Chen}
 \affiliation{Department of Applied Physics, School of
 Medical Imaging, Bengbu Medical University, Bengbu 233030, China.}

\author{Kai-Yuan Zhang}
 \affiliation{School of Physics and Wuhan National High Magnetic Field Center,
Huazhong University of Science and Technology, Wuhan 430074, People's Republic of China.}

\author{Xi Sun}
 \affiliation{School of Physics and Wuhan National High Magnetic Field Center,
Huazhong University of Science and Technology, Wuhan 430074, People's Republic of China.}

\author{Shu-Zheng Zhou}
 \affiliation{School of Physics and Wuhan National High Magnetic Field Center,
Huazhong University of Science and Technology, Wuhan 430074, People's Republic of China.}

\author{Hua-Hua Fu$^{*,}$}
\affiliation{School of Physics and Wuhan National High Magnetic Field Center,
Huazhong University of Science and Technology, Wuhan 430074, People's Republic of China.}
\affiliation{Institute for Quantum Science and Engineering, Huazhong University of Science and Technology, Wuhan, Hubei 430074, China.}
\date{November 25, 2025}
\begin{abstract}
\noindent
\noindent{\bfseries \textcolor[RGB]{31,51,128}{ABSTRACT:}} Recent progress in constructing supramolecular assemblies with hierarchical chirality offers new opportunities to investigate the chirality-induced spin selectivity (CISS) effect and its potential applications. In this work, we systematically examine the CISS effect in such multichiral systems by designing a class of multilayer helical architectures constructed of stacked and interfaced individual helical rings, each possessing well-defined local chirality. Through controlled interlayer twisting, a global helical handedness is further imposed, forming a multichiral tubular helix. Theoretical calculations reveal that these two distinct chiral hierarchies lead to several unprecedented CISS phenomena, such as enhanced spin polarization arising from cooperative dual chirality, along with the simultaneous emergence of transverse and longitudinal CISS signals. Moreover, interlayer torsional competition modulates the system's response to external fields. The dual-chiral geometry breaks the conventional symmetry of single helices, inducing an anomalous angular phase shift in magnetoresistance. Furthermore, Floquet analysis reveals that the interplay between local and global chirality enables controlled spin polarization switching under circularly polarized light. These findings provide a basic theoretical framework for studying the CISS in multichiral superstructures and establish design principles for coupled optical, magnetic, and spin manipulations, thereby facilitating the development of multichiral spintronic devices.

\end{abstract}
\maketitle

\ACSection{INTRODUCTION}
\noindent Chirality, which represents a fundamental break in spatial symmetry in which a structure cannot be superposed with its mirror image, plays a central role in the governance of stereospecific interactions among biological systems and in the generation of distinct optical responses under circularly polarized light \cite{Tang2010,Greenfield2021,Lininger2023}. A deeper quantum manifestation of chirality has recently emerged through the chirality-induced spin selectivity (CISS) effect, in which electron transmission through a chiral medium becomes spin-dependent \cite{Gohler2011}. This spin-filtering behavior offers a promising platform for developing room-temperature spintronic devices that may operate purely electrical means, frequently without reliance on ferromagnetic materials \cite{Bloom2024}. Mounting theoretical and experimental studies further point to an intrinsic connection between CISS and molecular electronic structure, with models suggesting a common physical origin with optical activity rooted in anisotropic polarizability \cite{Metzger2023}. Accordingly, CISS phenomena have been extensively documented in chiral molecular systems whose handedness arises from either local stereogenic elements or well-defined helical scaffolds. Notable examples span biological macromolecules such as DNA \cite{Gohler2011,Das2022,Xie2011,Zwang2016,Abendroth2017,Mishra2019} and $\alpha$-helical peptides \cite{Mishra2013,Mishra2020,Rahman2022,BenDor2017, Theiler2023,Aragones2025}, along with $\pi$-conjugated helicenes \cite{Kiran2016,Kettner2018,Rodriguez2022,Liang2022,Giaconi2023} and chiral organic-inorganic hybrid perovskites \cite{Lu2019,Lu2020,Wang2021,Dong2025}.                        

The search for the microscopic origins of CISS has led to the development of various theoretical models and inspired experimental studies across a wide spectrum of chiral materials \cite{1.1,1.2,1.3,Fransson2020, 1.3.1,Chiesa2024,Xu2024,Savi2025,1.5,1.6,1.7,1.9,1.12,1.12.1,1.13,1.14, Hoff2021,Naskar2023,ChenFu2023,ChenFu2024a,ChenFu2024b}. Early work largely revolved around molecular chirality originating from stereogenic centers. More recently, attention has expanded to include topological \cite{Zhang2023, Sun2025} and supramolecular \cite{ Mondal2021, Sang2023, Wang2024} chirality, where achiral building units are assembled into helical superstructures exhibiting collective handedness and efficient spin filtering. These advances have refined our understanding of the underlying physical mechanisms, suggesting that spin–orbit coupling (SOC), while essential, can produce substantial spin polarization through coherent accumulation of minor contributions, further enhanced by quantum interference \cite{Varela2016}. A central unresolved issue involves the role of momentum transfer in spin selectivity. In flexible biomolecules, electron–phonon coupling often facilitates vibration-assisted transport \cite{1.2,1.3,Fransson2020}; in contrast, rigid $\pi$-conjugated systems appear to employ a more direct mechanism involving electron–electron interactions between propagating electrons and the delocalized electron density of the molecular framework \cite{1.3.1, Chiesa2024,Xu2024,Savi2025}.

Although these studies highlight the significance of chirality across structural hierarchies, the majority have focused on systems governed by a single dominant chirality. In contrast, the integration of multiple and distinct forms of chirality, such as local chirality arising from molecular-scale twisting and global chirality encoded in supramolecular helices, into a single architecture and the resulting emergent physical phenomena, remain considerably less explored. Although such multi-chiral supramolecular systems have been experimentally realized \cite{Ikai2020, Yang2018, Hifsudheen2017,Hano2025}, the associated CISS effect has unfortunately not yet been systematically investigated. In particular, a coherent theoretical framework capable of describing the spin-selective phenomena arising from such dual chirality is still lacking. Therefore, elucidating the synergy and competition among hierarchically organized chirality is essential not only for understanding the interplay of chirality across scales, but also for unveiling the underlying mechanisms of chirality-driven spin control.

\begin{figure*}
\includegraphics[width=2\columnwidth]{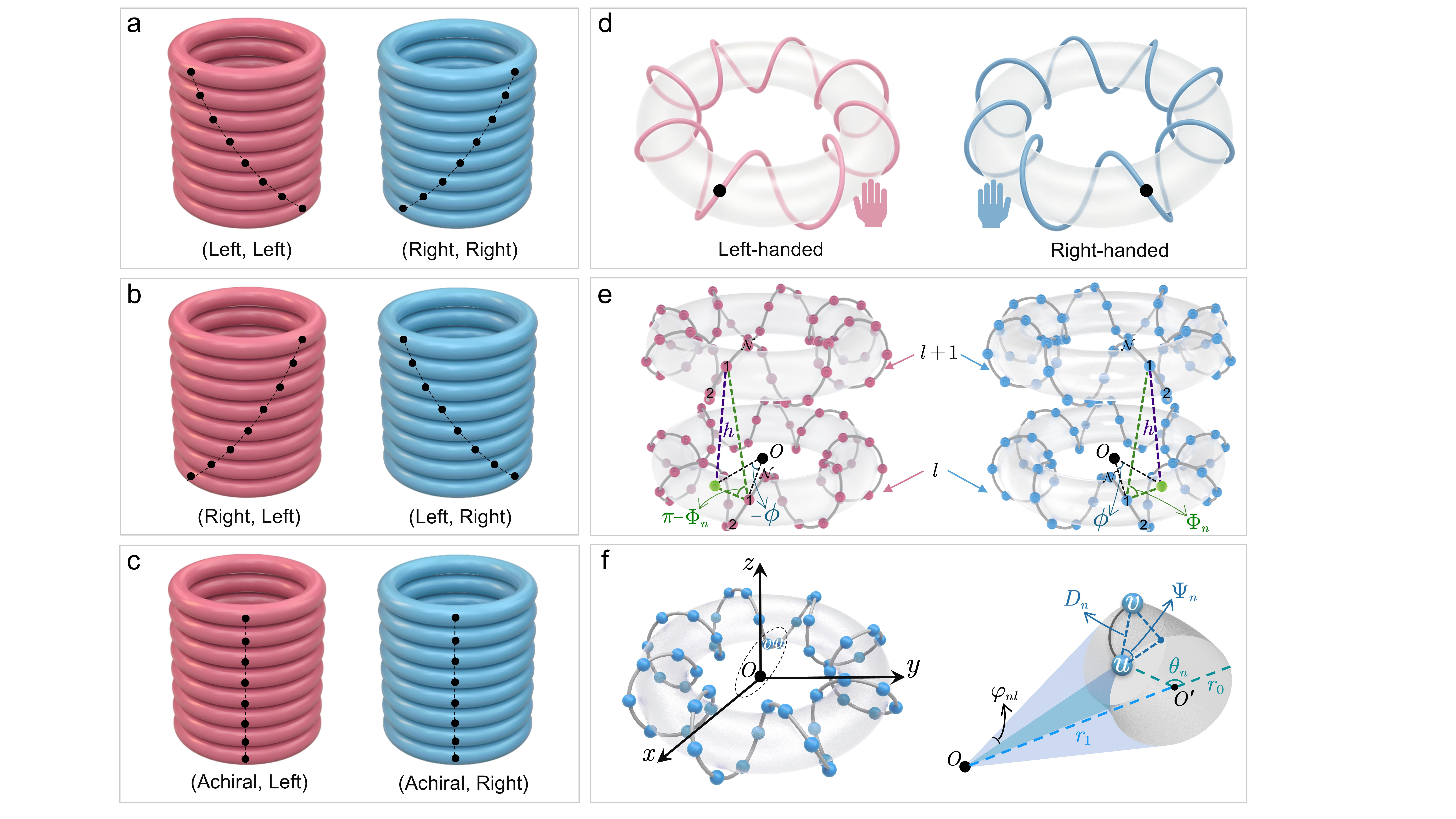}
\caption{Schematic of the MCTH model and its geometric parameters. (a) (Left, Left) configuration with left-handed molecules (pink) twisted clockwise between layers, and its mirror image (Right, Right) with right-handed molecules (blue) twisted counterclockwise. (b) Mixed-chirality configurations: (Right, Left), consisting of left-handed molecules stacked with an anticlockwise (right-handed) twist, and its enantiomer (Left, Right). (c) Achiral stacking configurations: (Achiral, Left) and (Achiral, Right), where left- or right-handed molecules are stacked vertically without inter-layer twist. (d) Elementary building blocks: left-handed (pink) and right-handed (blue) circular helical molecule. (e) Magnified side view of two adjacent layers ($l$ and $l+1$) from the (Left, Left) (pink) and (Right, Right) (blue) models, showing inter-layer geometric parameters: layer spacing $h$, twist angle $\phi$, and inter-molecular space angle $\Phi_n$, illustrated using the (Right, Right) model. (f) Intra-layer geometry of a single circular helical unit in a Cartesian frame (left), and detailed local parameterization (right), where $O$ ($O'$) and $r_1$ ($r_0$) denote the center and radius of the toroidal molecule and its cross-section plane, respectively; $\varphi_n$ and $\theta_n$ are the toroidal and poloidal angles; $\Psi_n$ and $D_n$ represent the space angle and interatomic distance.}
\label{fig1}
\end{figure*}

In this work, we present a theoretical investigation of the CISS effect in a multichiral supramolecular architecture composed of stacked helical rings with well-defined local handedness. This hierarchical design enables the coexistence and interaction of both molecular and supramolecular chirality, offering a tunable platform to examine how the dual‐chiral organization influences spin‐selective electron transport. Through systematic quantum transport modeling, we demonstrate that the interplay between local and global chirality gives rise to a pronounced CISS response, including the emergence of transverse spin polarization even in untwisted stacks of helical units, highlighting the essential role of structural hierarchy in spin filtering. Moreover, the magnitude of the spin polarization is found to scale with the number of layers and is modulated by the relative strength of intra‐ versus interlayer SOC, as well as by electrode contact geometry. Incorporating electron–phonon interactions into the model further reveals that spin polarization not only persists but can be enhanced with increasing temperature, maintaining robust performance above room temperature.

In the regime of magnetic control, the interlayer twisting introduces structural competition that fundamentally alters the angular response. Unlike conventional single-helical systems, the dual-chiral geometry breaks axial symmetry and tilts the effective spin-filtering axis, resulting in an anomalous angular phase shift in magnetoresistance. Moreover, Floquet analysis indicates that the hierarchical chirality interacts with optical helicity, breaking conventional enantiomeric symmetry and enabling all-optical switching of spin polarization. By establishing hierarchical chirality as a central design parameter, this theoretical study provides a foundation for multi‐chiral engineering in spintronic systems and reveals mechanisms for coupled control over optical, magnetic and spin degrees of freedom.\\

\ACSection{RESULTS AND DISCUSSION}
\noindent {\fontfamily{SourceSansPro-TLF}\fontseries{b}\selectfont MCTH Model and Circular Dichroism.}
To systematically explore spin-dependent transport hierarchical chiral systems, we introduce a class of multilayer helical architectures, constructed by stacking and intercoupling individual helical rings with well-defined local chirality. The ordered interlayer twisting further introduces a global helical handedness, resulting in what we term a multichiral tubular helix (MCTH). Each assembly consists of circular helical molecules as fundamental units, a closed-loop topology that offers key advantages over conventional linear helices. First, cyclic geometry eliminates end effects, conferring enhanced topological stability against thermal fluctuations and structural mismatch during stacking. Second, the ring structure supports coherent electron delocalization along clockwise and counterclockwise pathways, forming an intrinsic interferometer. Due to SOC, electrons traversing these paths accumulate spin-dependent Berry phases of opposite sign. Their subsequent quantum interference leads to constructive enhancement for one spin state and destructive cancellation for the other, a mechanism absent in linear transport, which enables substantially higher spin polarization.

Within this framework, we construct four distinct MCTH models organized as two enantiomeric pairs by systematically varying the inter-ring twist angle (Figures \ref{fig1}a and b). These are labeled as (Left, Left) $\&$ (Right, Right) and (Right, Left) $\&$ (Left, Right), where the first term denotes the global handedness of the MCTH and the second refers to the intrinsic helicity of the molecular rings. We also examine an achiral stacking arrangement without inter-ring twist, designated as (Achiral, Left) and (Achiral, Right) (Figure \ref{fig1}c). The electronic properties of the MCTH are described by the following effective tight-binding Hamiltonian 
$\mathcal{H}_\text{MCTH}=\mathcal{H}_{c}+\mathcal{H}_{\mathrm{soc}}, $
where $\mathcal{H}_{c}$ describes the kinetic and potential energy of the discrete lattice, and $\mathcal{H}_{\mathrm{soc}}$ incorporates SOC at both intramolecular and intermolecular levels. The first term, $\mathcal{H}_{c}$, is given by:
\begin{equation}
\begin{aligned}
\mathcal{H}_c 
= \sum_{n=1}^{\mathcal{N}} & \Bigg(
    \sum_{l=1}^{\mathcal{L}}
        \Big(
            \varepsilon_{nl} c_{nl}^{\dagger} c_{nl}
            + t_{nl,n+1l} c_{nl}^{\dagger} c_{n+1,l}
        \Big)
    \\
    &\quad + \sum_{l=1}^{\mathcal{L}-1}
        t_{nl,nl+1} c_{nl}^{\dagger} c_{n,l+1}
\Bigg)
+ \mathrm{H.c.},
\label{eq1}
\end{aligned}
\end{equation}
here, $c_{nl}^{\dagger}=(c_{nl \uparrow}^{\dagger}, c_{nl \downarrow}^{\dagger})$ is the creation operator and $\varepsilon_{nl}$ is the onsite energy of an electron at the site $n$ of the layer $l$, for a system of $\mathcal{L}$ layers with $\mathcal{N}$ lattice sites in every layer. The nearest-neighbor hopping integral, $t_{\xi\overline{\xi },\xi+1\overline{\xi }}$, with $\xi=n, \overline{\xi }=l$ for intra-molecular hopping and $\xi=l,\overline{\xi}=n$ for inter-molecular hopping, and is described by a Slater-Koster-type parameterization that explicitly depends on the distance and relative orientation of the two $p_z$ orbitals:
\begin{equation}
\begin{aligned}
t_{\xi\overline{\xi},\xi+1\overline{\xi}} 
&= -V_{pp\pi} 
\exp\!\left(-\frac{|\mathbf{R}_{\xi\overline{\xi},\xi+1\overline{\xi}}| - d_{\parallel}}{l_c}\right)
\frac{|\mathbf{R}_{\xi\overline{\xi},\xi+1\overline{\xi}} \cdot \mathbf{e}_{\parallel}|^{2}}
     {|\mathbf{R}_{\xi\overline{\xi},\xi+1\overline{\xi}}|^{2}} 
\\[6pt]
&\quad 
+ V_{pp\sigma} 
\exp\!\left(-\frac{|\mathbf{R}_{\xi\overline{\xi},\xi+1\overline{\xi}}| - d_{\perp}}{l_c}\right)
\frac{|\mathbf{R}_{\xi\overline{\xi},\xi+1\overline{\xi}} \cdot \mathbf{e}_{\perp}|^{2}}
     {|\mathbf{R}_{\xi\overline{\xi},\xi+1\overline{\xi}}|^{2}}.
\end{aligned}
\end{equation}
This term comprises two contributions arising from orbital overlaps of the $\pi$-type (in-plane) and the $\sigma$-type (out-of-plane), with interaction strengths governed by $V_{pp\pi}$ and $V_{pp\sigma}$, respectively. The spatial decay of these overlaps is modeled by an exponential dependence on the interatomic distance $|\mathbf{R}_{\xi,\xi+1}|$, scaled by the characteristic length $l_c$, where $d_{\parallel}$ and $d_{\perp}$ denote the reference bond lengths for intra- and inter-layer couplings. The geometric anisotropy of the electronic interaction is further captured by the squared direction cosines, which project the bond orientation $\mathbf{R}_{\xi,\xi+1}$ onto the in-plane ($\mathbf{e}_{\parallel}$) and out-of-plane ($\mathbf{e}_{\perp}$) axes.

\begin{figure*}
\includegraphics[width=1.95\columnwidth]{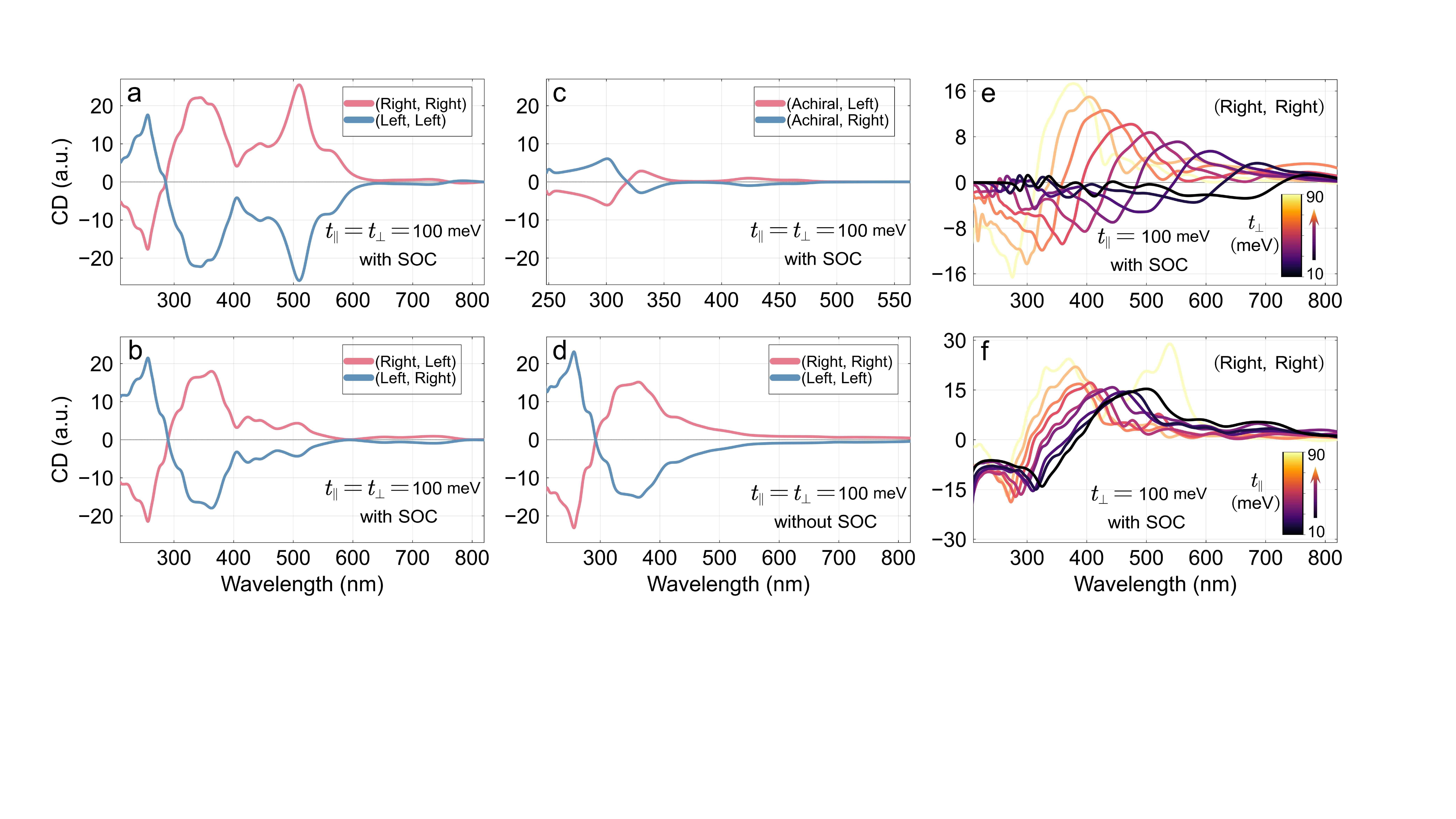}
\caption{Calculated CD spectra for MCTH models. (A) CD spectra of (Right, Right) and (Left, Left) configurations. (b) CD spectra of mixed-chirality (Right, Left) and (Left, Right) configurations. (c) CD spectra of achirally stacked (Achiral, Left) and (Achiral, Right) configurations. Spectra in (a-c) are calculated with SOC and fixed hopping parameters $t_{||}$=$t_{\perp}$=$100$ meV. (d) CD spectra of (Right, Right) and (Left, Left) models calculated without SOC, using the same hopping parameters. (e) Evolution of CD spectra for the (Right, Right) model (with SOC) as the inter-layer hopping ($t_{\perp}$) varies from 10 meV to 90 meV, while intra-layer hopping fixed at $t_{||} = 100$ meV. (f) Evolution of CD spectra for the (Right, Right) model (with SOC) as the intra-layer hopping ($t_{||}$) varies from 10 meV to 90 meV, while inter-layer hopping fixed at $t_{\perp} = 100$ meV.
}
\label{fig2}
\end{figure*}

The second term, $\mathcal{H}_{\mathrm{soc}}$, represents the chirality-induced geometric SOC written as
\begin{equation}
\begin{aligned}
\mathcal{H}_{\text{soc}} = &\sum_{n=1}^{\mathcal{N}} \Bigg[\sum_{l=1}^{\mathcal{L}}  {i\lambda _{\parallel}c_{nl}^{\dagger}\left( \sigma _{nl}^{\parallel}+\sigma _{n+1,l}^{\parallel} \right) c_{n+1,l}} \\
& + \sum_{l=1}^{\mathcal{L}-1} {i\lambda _{\bot}c_{nl}^{\dagger}\left( \sigma _{nl}^{\bot}+\sigma _{n,l+1}^{\bot} \right) c_{n,l+1}} + \mathrm{H}.\mathrm{c}. \Bigg],
\label{eq3}
\end{aligned}
\end{equation}
where $\lambda _{\parallel }$ and $\lambda _{\bot}$ denote the strength of intra- and inter-layer 
effective geometric SOC. 
\(\sigma_{n+1,l}^{\parallel}
= (\sin\!\theta_n\cos\!\varphi_{nl}\sin\!\varPsi_{n} -\sin\!\varphi_{nl}\cos\!\varPsi_{n})\sigma_x
+ (\sin\!\theta_n\sin\!\varphi_{nl}\sin\!\varPsi_{n} + \cos\!\varphi_{nl}\cos\!\varPsi_{n})\sigma_y
+ \cos\!\theta_n\sin\!\varPsi_{n}\sigma_z\), and \(\sigma_{n,l+1}^{\perp}
= \sin\!\varTheta_n(\sin\!\varphi_{nl}\sigma_x
- \cos\!\varphi_{nl}\sigma_y)
+ \cos\!\varTheta_n\cos\!\tfrac{\phi}{2}\sigma_z\),
with $\varphi_{nl}=(n-1)\Delta\varphi +l\phi$, $\Delta \varphi=2\pi/\mathcal{N}$, and $\theta _n=(n-1)\Delta \theta$; $\Delta \theta=2\pi /\mathcal{M}$ with $\phi$ the twist angle and $\mathcal{M}$ the total number of lattice sites per unit cell in each molecular layer. 
The intra-molecular space angle, $\varPsi_{n}$, is defined as $\varPsi_{n}=\mathrm{arc}\cos \left[ X_{n}/D_{n} \right]$, where $X_{n}=2 r_0 \sin \left[\Delta \theta/2\right]$ and this distance $D_{n}$ can be represented by an analytical approximation, derived by modeling the geometry between the site $n$ (with radius $\rho_n$) and the site $n+1$ (with radius $\rho_{n+1}$) as a planar isosceles trapezoid and applying Ptolemy's Theorem. This model yields the expression $
D_{n} = 2 \sqrt{\rho_n\rho_{n+1} \sin^2\left(\Delta\varphi/2\right) + r_0^2 \sin^2\left(\Delta\theta/2\right)}
$, where the radial length is $\rho_n = r_1 + r_0 \cos\theta_n$. Similarly, the inter-molecular space angle $\varTheta_n$ is given by $\varTheta_n=\mathrm{arc}\cos \left[ 2\rho_n\sin (\phi/2)/\left| \ell_n \right|\right]$, where the length of the bond is $\left| \ell_n \right|=\sqrt{h^2+(2\rho_n\sin(\phi/2))^2}$, with $h$ being the distance between layers.

We computed the circular dichroism (CD) spectra for each model to confirm the chiral characteristics of the designed hierarchical structures. The CD signal is defined as the difference in the absorption coefficients of circularly polarized light with left- and right-handedness \cite{Qiu2024,Apergi2023}. Within the tight-binding framework, the frequency-dependent absorption coefficient $\alpha_{\eta}(\omega) (\eta = L/R)$ is derived from the complex dielectric function $\epsilon_{\eta}(\omega) = \epsilon_{r,\eta}(\omega) + i\epsilon_{i,\eta}(\omega)$ as follows
\begin{equation}
\alpha_{\eta}(\omega)=\sqrt{2}\frac{\omega}{c}\sqrt{-\epsilon_{r,\eta}(\omega)+\sqrt{\epsilon_{r,\eta}^{2}(\omega)+\epsilon_{i,\eta}^{2}(\omega)}}
\end{equation}
The imaginary component of the dielectric function, $\epsilon_{i,\eta}(\omega)$, serves as the central quantity in this calculation. It is evaluated by adding all allowed electronic transitions between the occupied valence ($v$) and unoccupied conduction ($c$) bands:
\begin{equation}
\epsilon_{i,\eta}(\omega) \propto \frac{1}{\omega^2} \sum_{\mathbf{q},\mathbf{k}} |\langle c\mathbf{q}|\hat{\mathbf{u}}_{\eta}\cdot \mathbf{v}|v\mathbf{k}\rangle|^{2}\delta(E_{c}(\mathbf{q})-E_{v}(\mathbf{k})-\hbar\omega).
\end{equation}
A non-zero CD signal requires the transition matrix element $\langle c\mathbf{q}|\hat{\mathbf{e}}_{\eta}\cdot \mathbf{v}|v\mathbf{k}\rangle$ to be evaluated beyond the electric dipole approximation, explicitly incorporating electric quadrupole-magnetic dipole contributions. Complete implementation details are provided in the Supporting Information.

\begin{figure*}
\includegraphics[width=1.95\columnwidth]{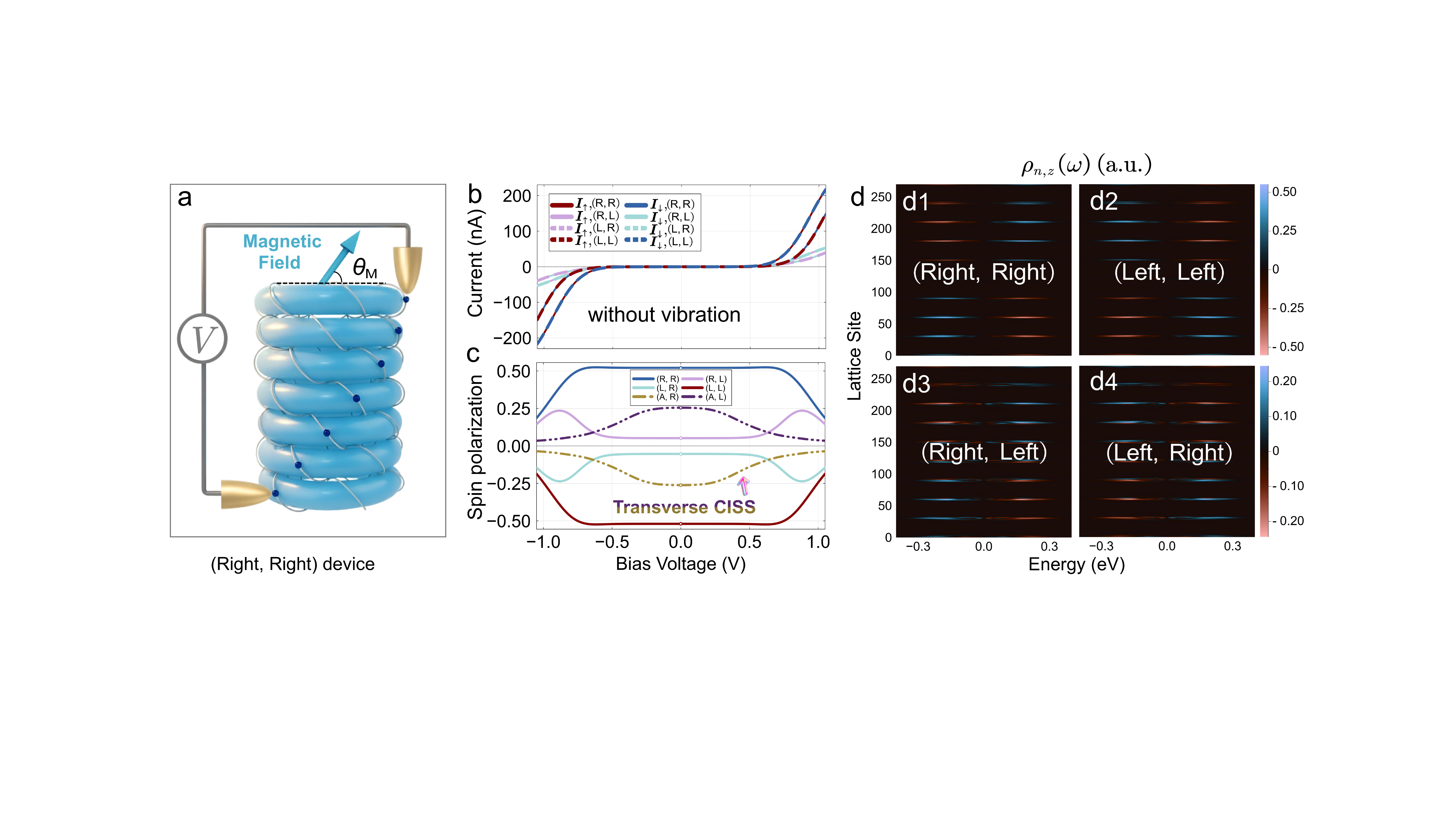}
\caption{Hierarchical CISS effect in MCTH models. (a) Schematic of the two-terminal device: the MCTH is connected to a non-magnetic top electrode and a ferromagnetic bottom electrode, whose magnetization direction is set by an external magnetic field at angle $\theta_M$ relative to the helical axis. (b) Current-voltage (I-V) characteristics under opposite magnetization directions. (c) Corresponding spin polarization $P_S$ as a function of bias voltage for for different chiral configurations, calculated without electron-phonon coupling. (d) Spin-projected local density of states $\rho_{s,n}(\omega)$ for (d1) (Right, Right), (d2) (Left, Left), (d3) (Right, Left), and (d4) (Left, Right) models.}
\label{fig3}
\end{figure*}

The calculated CD spectra exhibit distinct chiroptical signatures that correlate directly with the hierarchical chirality of the system (Figure \ref{fig2}). All configurations display oscillatory Cotton effects with multiple sign reversals across the absorption range. As expected, CD intensity diminishes at longer wavelengths, where photon energies drop below the electronic band gap. Enantiomeric pairs, both in synergetic (e.g., R/R vs. L/L, Figure \ref{fig2}a) and competitive (e.g., R/L vs. L/R, Figure \ref{fig2}b) configurations, produce strict mirror-image spectra, confirming chirality fidelity at both local and global levels. In particular, the CD amplitude of the twisted assemblies is substantially enhanced relative to that of achirally stacked reference systems (Figure \ref{fig2}c), demonstrating that superimposing a global supramolecular twist onto locally chiral units cooperatively amplifies the overall chiroptical response. To isolate the origin of this enhancement, calculations were performed without SOC (Figure \ref{fig2}d). The persistence of a strong antisymmetric CD profile confirms that the optical activity arises primarily from the geometric hierarchy and higher-order multipole contributions, rather than SOC effects.

Further insight into hierarchical chiral coupling is provided by examining the spectral dependence on electronic coupling parameters (Figures \ref{fig2}e and f). Variation of the hopping integrals shifts the Cotton effect profiles, reflecting modulation of the transition energies and band structure. Importantly, CD intensity shows markedly stronger dependence on interlayer coupling ($t_{\perp}$) than on intralayer coupling ($t_{||}$), indicating that global supramolecular packing, rather than local molecular connectivity, dominates chiroptical output. This pronounced sensitivity to electronic communication between layers underscores the critical role of the higher-order chiral architecture in modulating intrinsic molecular CD signals, thus establishing a clear structure–property relationship governed by dual chirality in the MCTH assemblies.\\

\noindent {\fontfamily{SourceSansPro-TLF}\fontseries{b}\selectfont Hierarchical Chirality-Induced Spin Selectivity.}
Spin transport properties are simulated using a two-terminal device geometry (Figure \ref{fig3}a), where the MCTH structure bridges a nonmagnetic top electrode and a ferromagnetic bottom electrode. The magnetization direction of the bottom electrode, determined by an external magnetic field, is defined by the angle $\theta_M$ relative to the assembly’s helical axis. We first examine the case with magnetization parallel to the transport direction ($\theta_M$=$\pi/2$). The total Hamiltonian is given by 
\begin{equation}
\mathcal{H} = \mathcal{H}_{\text{MCTH}} + \mathcal{H}_{\text{EP}} + \mathcal{H}_{\text{M}} + \mathcal{H}_{\text{N}} + \mathcal{H}_{\text{D}},
\end{equation}
where $\mathcal{H}_{\text{MCTH}}$ describes the electronic structure of the multichiral assembly, $\mathcal{H}_{\text{M}}$ and $\mathcal{H}_{\text{N}}$ represent the magnetic and nonmagnetic electrodes, and $\mathcal{H}_{\text{D}}$ accounts for the electrode-molecule coupling.

To capture the effect of lattice vibrations in these flexible molecular stacks, we explicitly include electron–phonon interactions via $\mathcal{H}_{\text{EP}}$. Beyond conventional charge scattering, phonons in chiral SOC systems actively contribute to spin selectivity by modulating both the electrostatic potential and the SOC strength. This results in a spin-dependent electron–phonon coupling that effectively lifts spin degeneracy through a vibrationally induced exchange splitting. The electron–phonon term is partitioned as $\mathcal{H}_{\text{EP}}$=$\mathcal{H}_{\text{ph}}$+$\mathcal{H}_{\text{e-ph}}$, where $\mathcal{H}_{\text{ph}}$=$\sum_{\nu} \hbar\omega_{\nu} a_{\nu}^{\dagger} a_{\nu}$ describes free phonons with mode frequency $\omega_{\nu}$ and operators $a_{\nu}^{\dagger}$ ($a_{\nu}$). The coupling term $\mathcal{H}_{\text{e–ph}}$ captures how lattice vibrations modulate both electron hopping and SOC, as described in the following
\begin{equation}
\begin{aligned}
\mathcal{H}_{\text{e-ph}} &= \sum_{n=1}^{\mathcal{N}}  \Bigg[
    \sum_{l=1}^{\mathcal{L}}
        \left( c_{nl}^{\dagger} c_{n+1,l} \sum_{\nu} t_{n,n+1}^{\nu} (a_{\nu} + a_{\nu}^{\dagger}) \right)
    \\
    \quad &+ \sum_{l=1}^{\mathcal{L}-1}
        \left( c_{nl}^{\dagger} c_{n,l+1} \sum_{\nu} t_{l,l+1}^{\nu} (a_{\nu} + a_{\nu}^{\dagger}) \right)
    \\
    \quad &+ \sum_{l=1}^{\mathcal{L}}
        \left( i c_{nl}^{\dagger}\left( \sigma _{nl}^{\parallel}+\sigma _{n+1,l}^{\parallel} \right) c_{n+1,l} \sum_{\nu} \lambda_{\parallel}^{\nu} (a_{\nu} + a_{\nu}^{\dagger}) \right)
    \\
    \quad &+ \sum_{l=1}^{\mathcal{L}-1}
        \left( i c_{nl}^{\dagger}\left( \sigma _{nl}^{\bot}+\sigma _{n,l+1}^{\bot} \right) c_{n,l+1} \sum_{\nu} \lambda_{\perp}^{\nu} (a_{\nu} + a_{\nu}^{\dagger}) \right)
\Bigg] \\
&+ \mathrm{H.c.}
\label{eq_eph}
\end{aligned}
\end{equation}
Here, $t_{n,n+1}^{\nu}$ and $t_{l,l+1}^{\nu}$ are the coupling constants between the phonon mode $\nu$ and the intra- and inter-molecular hopping, respectively. Similarly, $\lambda_{\parallel}^{\nu}$ and $\lambda_{\perp}^{\nu}$ are the coupling constants that describe the modulation of the intra- and inter-molecular SOC terms by the phonon mode $\nu$. The third and fourth terms $\mathcal{H}_{\text{N}}$ and $\mathcal{H}_{\text{M}}$ model the non-magnetic upper and magnetic lower electrodes, respectively, which are treated as semi-infinite electron reservoirs. Their Hamiltonians and coupling to the central assembly are expressed as: $\mathcal{H}_{\text{N}} = \sum_{k,\sigma} \epsilon_{Nk} b_{k\sigma}^{\dagger} b_{k\sigma} + \sum_{k,\sigma, i \in N} (\tau_N b_{k\sigma}^{\dagger} c_{i\sigma} + \text{H.c.}),  
\mathcal{H}_{\text{M}} = \sum_{k,\sigma} \epsilon_{Mk\sigma} d_{k\sigma}^{\dagger} d_{k\sigma} + \sum_{k,\sigma, i \in M} (\tau_M d_{k\sigma}^{\dagger} c_{i\sigma} + \text{H.c.})$.
Here, $b_{k\sigma}^{\dagger}$ ($d_{k\sigma}^{\dagger}$) creates an electron with momentum $k$ and spin $\sigma$ in the non-magnetic (magnetic) lead with energy $\epsilon_{Nk}$ ($\epsilon_{Mk\sigma}$). The coupling strength between the leads and the specific sites $i$ in the assembly is given by $\tau_N$ and $\tau_M$.
The final term, $\mathcal{H}_{\text{D}}$, accounts for the mechanisms of dephasing through electron leakage. This process is modeled by coupling every lattice site in the supramolecule with an independent virtual absorbing channel:
$\mathcal{H}_{\text{D}} = \sum_{n,l,j} (\epsilon_{Dj} f_{nlj}^{\dagger} f_{nlj} + \tau_D f_{nlj}^{\dagger} c_{nl} + \text{H.c.})$, where $f_{nlj}^{\dagger}$ creates an electron in the dephasing channel connected to the site $(n,l)$ of the main helix. This approach introduces a localized and complex self-energy at each position, thereby disrupting the system's unitarity.

\begin{figure*}
\includegraphics[width=1.95\columnwidth]{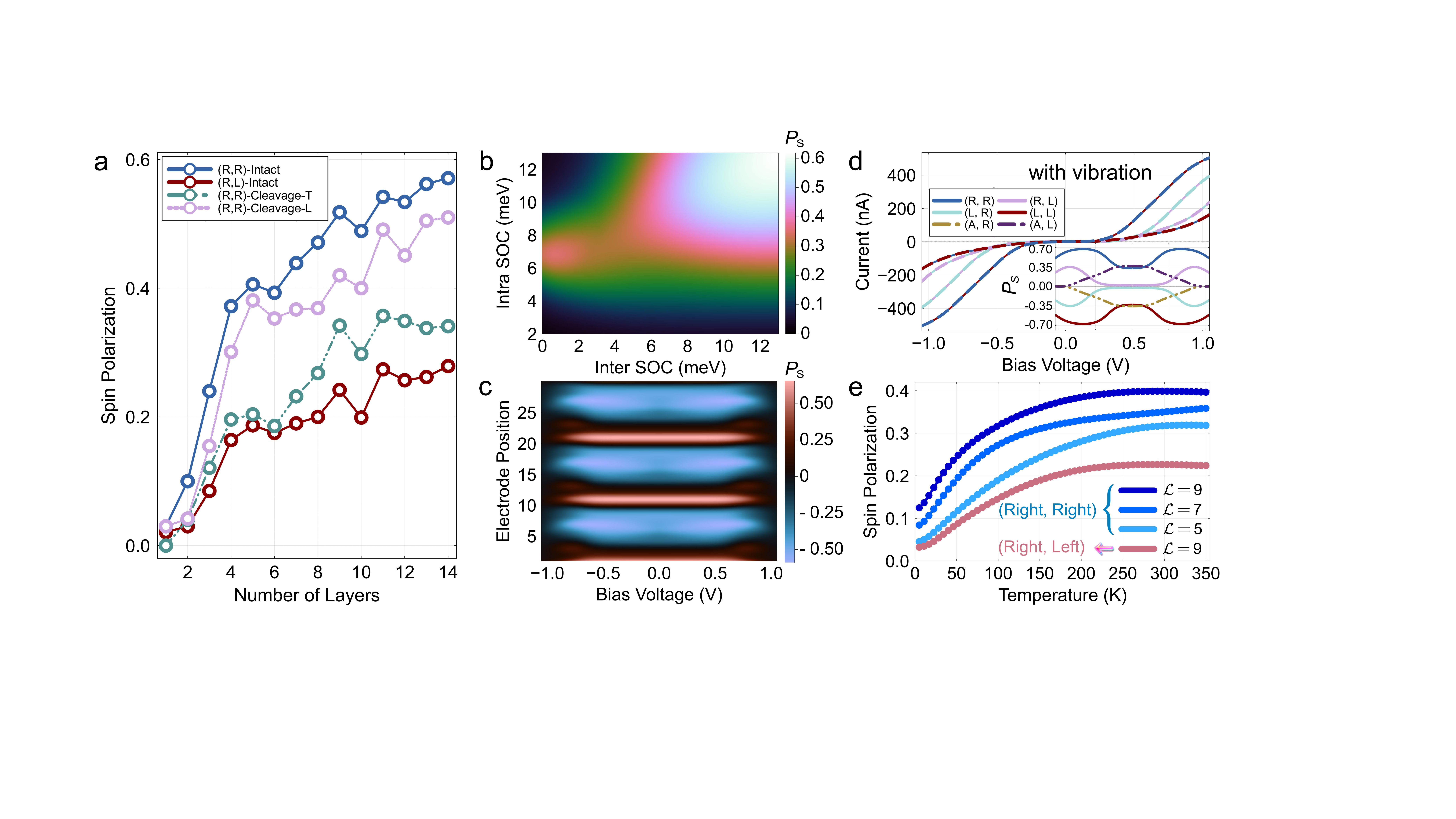}
\caption{Parametric modulation of the Hierarchical CISS effect in MCTHs. (a) Spin polarization as a function of the number of layers for intact and cleaved assemblies. (b) 2D map of spin polarization versus intra- and inter-layer SOC strengths. (c) 2D map of spin polarization as a function of bias voltage and top electrode contact position. (d) I-V characteristics with inset showing spin polarization, both calculated inlcuding electron-phonon coupling. (e) Temperature dependence of spin polarization for assemblies with varying numbers of layers.}
\label{fig4}
\end{figure*}

The spin-dependent electron transport properties of dual-chiral systems were investigated by simulating current-voltage (I–V) characteristics for different chiral configurations in the absence of electron-phonon coupling. As shown in Figure \ref{fig3}b, the current was evaluated for the opposite magnetization directions of the ferromagnetic electrode. To better resolve the current characteristics without altering the spin polarization behavior, the electrode coupling was moderately increased in competitive configurations. From these I–V curves, the spin polarization ($P_S$) was derived as a function of the bias voltage (Figure \ref{fig3}c). The results indicate that synergetic configurations, where local and global handedness align, exhibit significantly higher spin filtering efficiency than competitive ones, in which these chiral scales oppose each other. Enantiomeric pairs consistently yield antisymmetric spin polarization profiles, confirming the chirality-spin correspondence across hierarchical levels. Given that this spin selectivity arises from the concerted action of molecular helicity and supramolecular packing chirality, we refer to it as the hierarchical CISS effect.

Notably, considerable spin polarization is maintained even at zero interlayer twist angle. Since electron injection occurs perpendicular to the helical axis, this behavior represents a transverse CISS effect (Figure \ref{fig3}c). To gain insight into the electronic origin of these transport features, the spin-projected local density of states, $\rho _{n,z\left( \omega \right)}$, was analyzed with zero bias (Figure \ref{fig3}d). The synergetic models (panels d1, d2) exhibit a substantially larger magnitude of spin-asymmetric density near the Fermi level than the competitive models (d3, d4). Mirror-image spin textures are observed for enantiomeric configurations, which is consistent with the transport measurements. This intrinsic spin polarization of the electronic structure directly underlies the observed spin-polarized current and the tunable spin filtering behavior shown in Figure \ref{fig3}b,c, demonstrating that the interplay between local and global chirality governs both the magnitude and symmetry of spin transport in multichiral assemblies.

\begin{figure*}
\includegraphics[width=2.0\columnwidth]{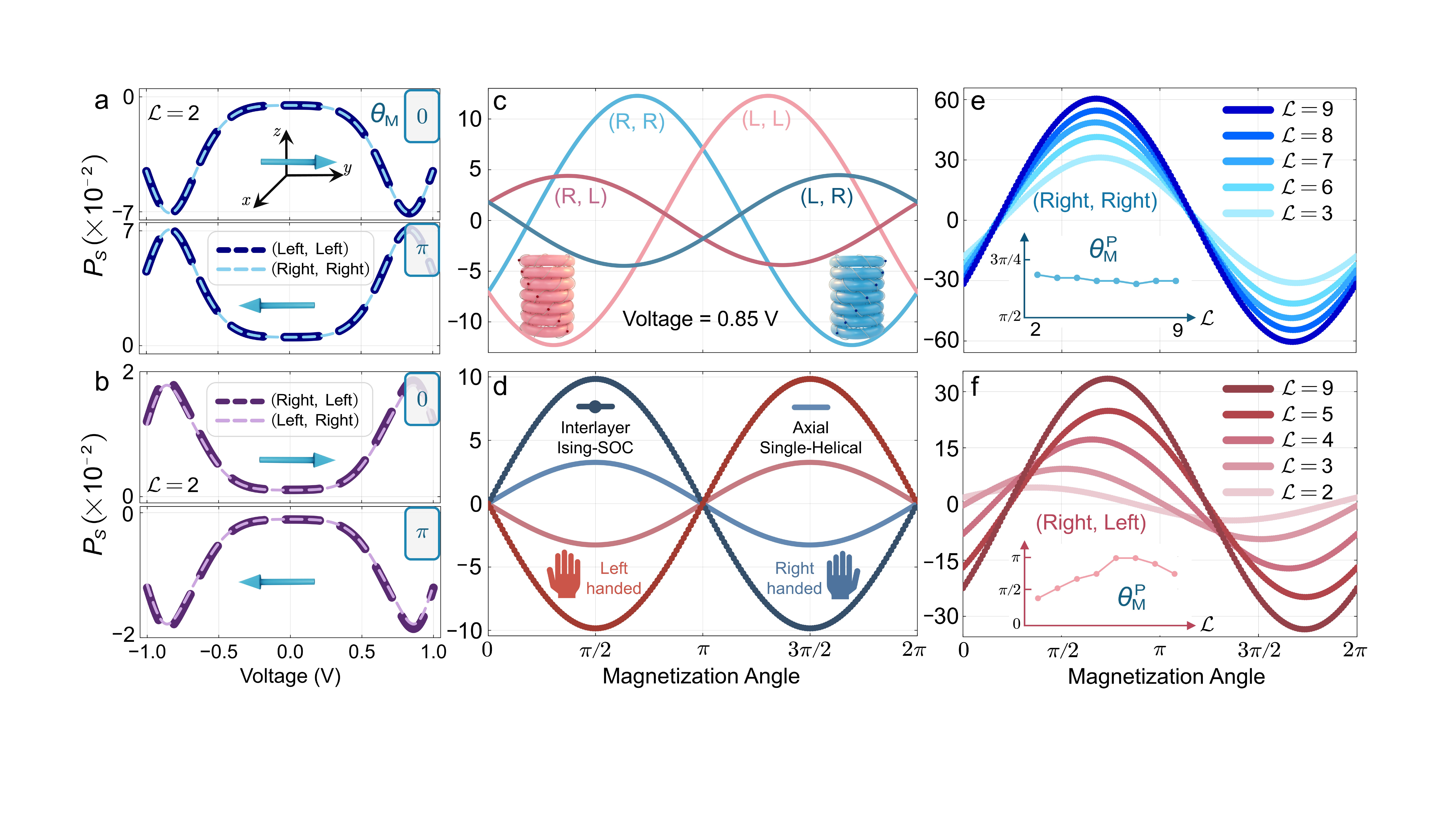}
\caption{Magnetic control of Hierarchical CISS. (a, b) Spin polarization $P_S$ as a function of bias voltage at magnetization angles $\theta_M = 0$ and $\pi$ ($\mathcal{L}=2$) for (a) the (Left, Left) and (Right, Right) configurations and (b) the (Right, Left) and (Left, Right) configurations. (c) Angular dependence of $P_S$ on $\theta_M$ for all four chiral configurations at a fixed bias of 0.85 V. (d) Comparison of $P_S$ vs. $\theta_M$ for left-handed (red) and right-handed (blue) configurations calculated using the interlayer Ising-SOC model (solid lines with markers) and the axial single-helical model (solid lines). (e, f) Evolution of the angular dependence of $P_S$ with increasing number of layers ($\mathcal{L}$) for (e) the (Right, Right) and (f) the (Right, Left) configurations. Insets: The peak magnetization angle $\theta_M^P$ as a function of $\mathcal{L}$.}
\label{fig5}
\end{figure*}

To evaluate the robustness of the hierarchical CISS effect, we systematically analyzed its dependence on key structural and electronic parameters. The spin polarization exhibits a monotonic increase with the number of layers in representative models, indicating that the spin filtering is enhanced coherently as the system extends along the stacking direction. In single-layer systems, both left- and right-handed molecular units show limited polarization ($\sim$$4\%$), with the spin quantization axis oriented perpendicular to the molecular helical axis. The marked amplification of spin polarization upon stacking and twisting highlights the critical role of the hierarchical chiral design, where the cooperative interplay between local helicity and global chiral packing governs the spin selectivity.

The multi-path quantum transport characteristics inherent to the dual-chiral geometry are further verified through controlled defect simulations. Deliberate disruption of intra-layer (transverse) or inter-layer (longitudinal) electronic coupling significantly suppresses spin polarization (Figure \ref{fig4}a), confirming that coherent electron transmission along both pathways is essential for the observed Hierarchical CISS effect. Parameter sensitivity analysis reveals that spin filtering is more strongly dependent on intra-molecular SOC than on inter-molecular SOC (Figure \ref{fig4}b), underscoring the primary contribution of local molecular chirality in establishing the spin-selective potential landscape. Furthermore, the spin response is highly sensitive to the electrode contact geometry. With the bottom electrode fixed at the first site of each unit cell, varying the top electrode contact position robustly modulates both the magnitude and sign of spin polarization (Figure \ref{fig4}c), illustrating how external probing conditions can selectively access distinct chiral interference patterns arising from the hierarchical structure.

In particular, incorporating electron–phonon interactions leads to a marked enhancement of spin polarization. Even with SOC strengths typical of organic systems \cite{Kuemmeth2008,Steele2013,Huertas2006}, the spin polarization increases systematically with both the number of layers and the temperature (Figure \ref{fig4}d,e). Throughout the entire temperature range studied, the synergetic chiral configuration consistently exhibits superior spin-filtering performance relative to the competitive arrangement (Figure \ref{fig4}e). This positive temperature dependence signifies a vibrationally assisted spin-filtering mechanism in which thermal fluctuations promote spin-selective transport, thereby enabling substantial spin polarization at room temperature while preserving the distinct characteristics of the dual-chiral interplay.\\

\noindent {\fontfamily{SourceSansPro-TLF}\fontseries{b}\selectfont Magnetic Field Control of Hierarchical CISS.}
Active control of spin transport through external fields represents a key requirement for practical spintronic applications. We first investigate the magnetic field orientation dependence of spin polarization using a bilayer system ($\mathcal{L}=2$) with enantiomeric pairs in dual-chiral configurations. The bottom electrode is modeled as a soft-to-medium ferromagnet (such as Ni or Co), where an applied field of several tens of millitesla sets the magnetization direction $\theta_M$ relative to the system's helical axis. Direct Zeeman effects on the molecular framework are neglected here.

Spin polarization as a function of bias voltage is evaluated at the principal magnetization angles (Figure \ref{fig5}a,b). The reverse of the magnetization direction consistently inverts the sign of $P_s$, confirming the robust magnetic control over spin transport. The continuous angular dependence (Figure \ref{fig5}c) further reveals fundamental constraints of the symmetry that governs the system. A strict antisymmetric enantiomeric relationship, $P_s(-\chi, \theta_M)$=$-P_s(\chi, \theta_M)$, is maintained only along the longitudinal axis ($\theta_M$=$\pm\pi/2$). At intermediate angles, this symmetry is broken due to competing influenced of local molecular and global suprramoelcular chirality. Furthermore, the spin response adheres to time-reversal symmetry, $P_s(\chi, \theta_M + \pi)$=$-P_s(\chi, \theta_M)$, and exhibits mirror symmetry, $P_s(\chi, -\theta_M)$=$P_s(-\chi, \theta_M)$. The latter symmetry leads to vanishing antisymmetric contributions in transverse orientations ($\theta_M=0, \pi$), resulting in identical spin polarization for opposite enantiomers (Figure \ref{fig5}a,b). These symmetry properties demonstrate how the dual-chiral architecture creates distinct spin-filtering behaviors that can be selectively accessed through magnetic field orientation.

Analysis of the angular dependence of spin polarization (Figure \ref{fig5}c) reveals a distinct behavior of the MCTH system compared to conventional single-helical architectures. Unlike previously reported molecular assemblies where magnetoresistance peaks align parallel to the helical axis ($\theta_M$=$\pi/2$) \cite{Das2024}, the MCTH model exhibits polarization extrema shifted to intermediate angles. This angular shift originates from the symmetry-breaking nature of the dual-chiral geometry, where the interplay between local molecular helicity and global supramolecular twisting modifies the effective spin-filtering axis. The origin of this anomalous angular response can be further elucidated by comparing it with two reference models of high-symmetry (Figure \ref{fig5}d). In the axial single-helical model, constructed by restricting electronic hopping to longitudinal pathways between the corresponding lattice sites, the maximum polarization is restored to $\theta_M$=$\pi/2$. Similarly, the interlayer Ising-type SOC model, which eliminates both interlayer transverse spin-flip processes and intralayer SOC contributions, also recovers the conventional angular dependence with peak polarization at $\theta_M$=$\pi/2$. These comparative results confirm that the angular shift observed in the MCTH system arises specifically from the cooperative interplay between intra-layer chiral transport and inter-layer spin-dependent coupling, demonstrating how hierarchical chirality fundamentally modifies the magnetic response of chiral spintronic systems.

\begin{figure*}
\includegraphics[width=1.95\columnwidth]{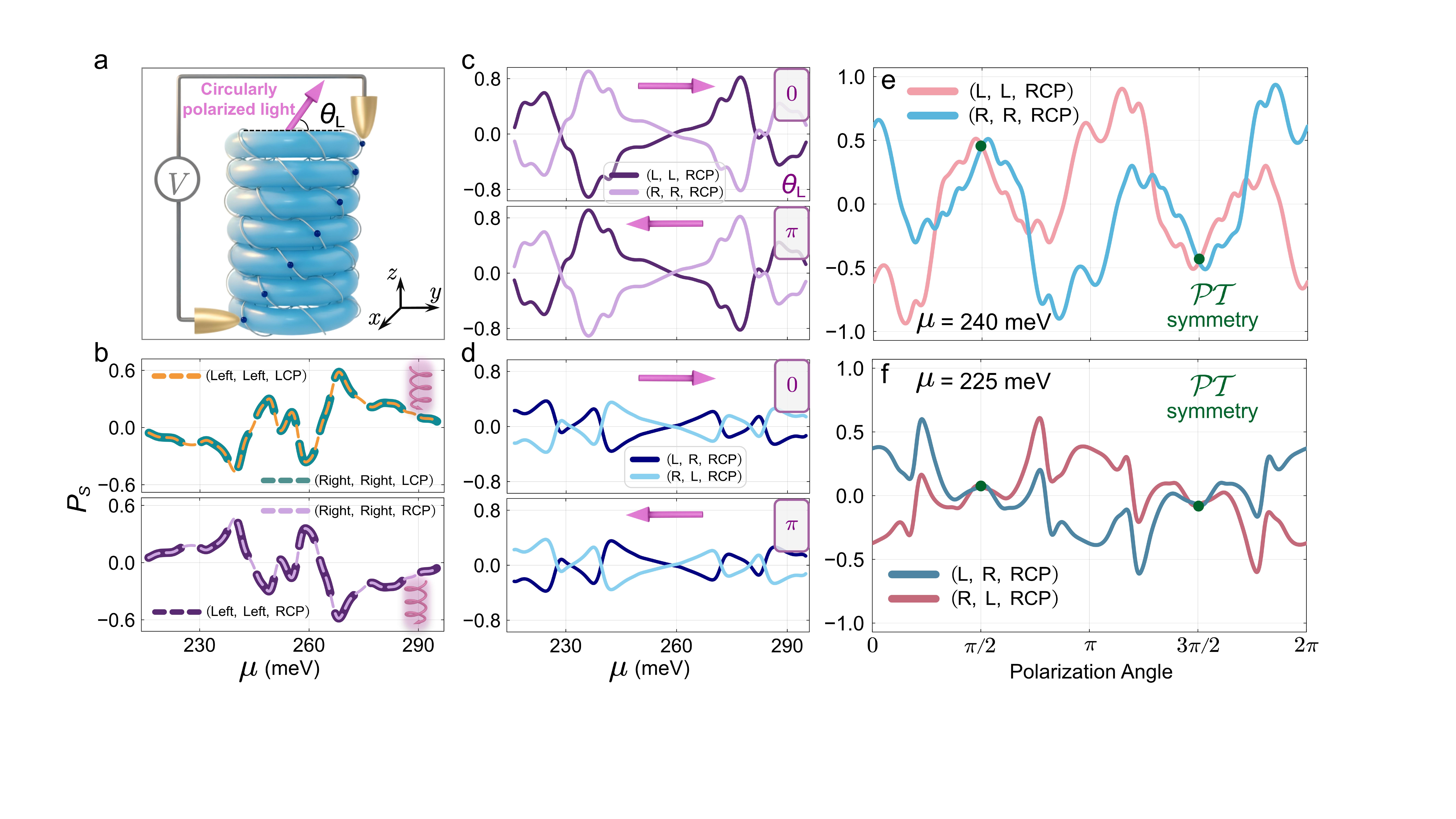}
\caption{Optical Manipulation of Hierarchical CISS. (a) Schematic of the device under circularly polarized light illumination with incidence angle $\theta_L$. (b) Spin polarization $P_S$ as a function of chemical potential $\mu$ for the synergetic (Left, Left) and (Right, Right) configurations under LCP and RCP light at $\theta_L = \pi/2$. (c, d) Comparison of $P_S$ vs. $\mu$ at $\theta_L = 0$ and $\pi$ for (c) the synergetic [(Left, Left), (Right, Right)] configurations and (d) the competitive [(Right, Left), (Left, Right)] configurations. (e, f) Angular dependence of $P_S$ on the light polarization angle $\theta_L$ at fixed chemical potentials for (e) the synergetic and (f) the competitive enantiomeric pairs. The green dots indicate points of $\mathcal{PT}$ symmetry where $P_s(\chi) = P_s(-\chi)$.}
\label{fig6}
\end{figure*}

The shifted maximum in spin polarization observed in the full MCTH models stems from symmetry breaking induced by a combination of intra-layer molecular helicity and inter-layer supramolecular twisting. In the axial single-helical reference model, high structural symmetry enables multi-path quantum interference that suppresses transverse SOC components, resulting in a spin-filtering axis aligned with the helical axis ($z$-direction). By contrast, the MCTH architecture involves sequential inter-layer electron hopping through a series of non-collinear SOC fields. This sequential process, governed by dual chirality, preserves transverse spin components and collectivity tilts the effective spin-filtering axis away from the $z$-direction. Consequently, maximum spin polarization occurs when the electrode magnetization aligns optimally with this tilted axis. It should be stressed that the layer-dependent evolution of spin polarization further supports this picture (Figure \ref{fig5}e,f). In the synergetic (Right, Right) configuration, a stable angular shift emerges and saturates as the number of layers increases. In contrast, the competitive (Right, Left) arrangement shows a continuously evolving optimal magnetization angle with system size (insets). These distinct behaviors highlight that the effective spin-filtering direction is not a local property but a global characteristic arising from the cumulative interplay between local and global chirality.

The observed angular phase shift in magnetoresistance carries important implications for the experimental characterization of chiral spintronic systems. Conventional analysis often employs an angular dependence of the form $\cos^2(\tilde{\theta})$ or $\sin^2(\tilde{\theta})$ (with $\tilde{\theta} = \theta_M + \pi/2$), which assumes a spin-filtering axis aligned with the high-symmetry direction of a single helix \cite{Das2024, McGuire1975, Zou2016, Nakayama2013}. In multi-layered chiral architectures, such as the MCTH system, this description becomes inadequate. The interlayer twist breaks the structural symmetry and tilts the effective spin-filtering axis, leading to an anomalous phase shift that cannot be captured by the standard model. This shift differs fundamentally from the dissipation-induced phase corrections proposed in other contexts \cite{Das2024}. Rather than arising from electron-phonon coupling, it originates from the structural desymmetrization imposed by the dual-chiral geometry, where local helicity and global twisting cooperatively lower the symmetry of the spin transport landscape. Consequently, the angular magnetoresistance line shape serves as a sensitive probe of both spin-dependent transport and structural symmetry. These findings motivate a generalized angular dependence of the form $\cos^2(\tilde{\theta}-\phi_0)$, where the phase offset $\phi_0$, often assumed to be zero, quantifies the structural desymmetrization inherent in hierarchical chiral materials.

Moreover, identifying this tilted spin-filtering axis establishes a principle of vectorial optimization for spin injection. Maximizing spin polarization efficiency, magnetoresistance, or charge current requires aligning the electrode magnetization with this effective axis rather than the geometric helical direction. This insight provides a concrete design rule for enhancing performance in devices based on multi-scale chiral molecular systems.\\

\noindent {\fontfamily{SourceSansPro-TLF}\fontseries{b}\selectfont All-Optical Control of Hierarchical CISS.}
The manipulation of spin-dependent electron transport via ultra-fast optical fields offers a promising route toward high-speed spintronic operation beyond conventional magnetic switching \cite{Lalieu2019,Dabrowski2022,Lambert2014,Nemec2018}. To examine this possibility in multichiral systems, we analyze the dynamics of electrons under circularly polarized irradiation using a Floquet engineering approach \cite{Wang2013,Kohler2005,Oka2009,McIver2020,Merboldt2025}. Within this framework, light-matter interaction is incorporated via the Peierls substitution, leading to a time-periodic Hamiltonian. Floquet theory then yields an effective static Hamiltonian whose quasi-energy spectrum governs spin-resolved transport in a two-terminal setup with magnetic components or dephasing. 

The Floquet-renormalized intra-layer ($\parallel$) and inter-layer ($\bot$) hoppling and SOC ($\lambda$) parameters are given by:
\begin{equation}
t(\lambda)_{\parallel(\bot)}^{(m)} = t(\lambda)_{\parallel(\bot)} i^m e^{im\delta_{\parallel(\bot)}(\theta_L)} \mathcal{J}_m\left(\mathcal{A}_{\parallel(\bot)}(\theta_L)\right),
\end{equation}
where $m$ denotes the net number of photons absorbed or emitted during tunneling and $\mathcal{J}_m$ denotes the $m$th-order spherical Bessel function. The effective coupling strength $\mathcal{A}_{\parallel(\bot)}(\theta_L)$=$eA_0\varXi_{\parallel(\bot)}(\theta_L)/\hbar$ depends on the amplitude of the vector potential $A_0$ and a geometric factor is given as 
\begin{equation}
\varXi_{\parallel(\bot)}(\theta_L) = \sqrt{\Delta x^2 + (\Delta y_{\parallel(\bot)}\sin\theta_L - \Delta z_{\parallel(\bot)}\cos\theta_L)^2}. 
\end{equation}
Then, the acquired tunneling phase is given as 
\begin{equation}
\delta_{\parallel(\bot)}(\theta_L) = \arctan\left(\eta\frac{\Delta y_{\parallel(\bot)}\sin\theta_L - \Delta z_{\parallel(\bot)}\cos\theta_L}{\Delta x_{\parallel(\bot)}}\right),
\end{equation}
which varies with the polarization angle $\theta_L$ (between the polarization plane and the helical axis) and the light polarity $\eta$=$\pm 1$ for the left/right circular polarization.

Under longitudinal illumination (light co-propagating with electrons along the helical axis), the (Right, Right) and (Left, Left) enantiomers exhibit nearly identical spin polarization (Figure \ref{fig6}b). This deviation from conventional antisymmetric enantio-response, where $P_s(\chi) = P_s(-\chi)$, reflects the underlying $\mathcal{PT}$ symmetry of the irradiated system. The Floquet analysis further confirms that optical spin control remains governed by the cooperation and competition between molecular and supramolecular chirality. As shown in Figures \ref{fig6}c and d, synergetic configurations produce significantly stronger spin polarization than competitive ones, indicating that the Floquet-dressed electronic states retain sensitivity to the structural hierarchy.

Angular sweeps of the polarization plane (Figures \ref{fig6}e,f) reveal systematic transformation rules: at $\theta_L = 0$ and $\pi$, the response is antisymmetric under enantiomer exchange, $P_s(-\chi, \theta_L)$=$-P_s(\chi, \theta_L)$. The system also obeys the time-reversal symmetry $P_s(\chi, \theta_L + \pi)$=$-P_s(\chi, \theta_L)$, and mirror transformation rule $P_s(\chi, -\theta_L)$=$-P_s(-\chi, \theta_L)$. A complete analysis of the $\mathcal{PT}$ symmetry linking optical helicity to hierarchical chirality, along with full transformation rules for hierarchical CISS under multi-field control, is provided in the Supporting Information.\\

\noindent {\fontfamily{SourceSansPro-TLF}\fontseries{b}\selectfont Experimental Corroboration of the Hierarchical CISS.} The synergistic and competitive interplay between local molecular and global supramolecular chirality, as revealed in our model, finds strong support in recent experimental studies of dual-chiral systems. In particular, our theoretical predictions align with observations from silica-helicene hybrid assemblies \cite{Hano2025}, where two key correspondences emerge: the global supramolecular handedness determines the overall sign of the chiroptical response, while structural mismatch in competitively chiral configurations induces measurable spectral shifts—consistent with electronic density redistribution and optical transition renormalization due to chiral incompatibility. These experimental observations align well with our theoretical results.

This interpretation is further supported by recent mechanical studies of macroscopic helical polymers, which present a compelling macroscopic parallel to the Hierarchical CISS phenomenon. In such systems, local molecular chirality governs the formation of the global helical architecture through a sequential process of chirality transfer and amplification across structural hierarchies \cite{Yang2018}. This mechanism is conceptually analogous to the cooperative reinforcement of spin polarization observed in our multi-layered theoretical models, where constructive interference between local and global chiral potentials leads to a cumulative enhancement of the effect.

From a functional perspective, a direct parallel exists in the system's operational principle: a condition of chiral compatibility between different structural levels facilitates efficient and ordered macroscopic actuation, mirroring the robust spin-selective transport enabled by aligned chiral fields in the synergetic MCTH system. Conversely, a mismatch in chirality results in disordered deformation at the macroscopic scale, paralleling the suppressed spin selectivity arising from competing structural motifs in the electronic transport context. These experimental observations from macroscopic polymer systems provide substantive and independent validation of our theoretical framework. They collectively demonstrate that cooperative coupling across distinct chiral hierarchies constitutes a universal organizational principle, which governs diverse material properties ranging from macroscopic mechanical response to quantum-mechanical spin transport in multi-chiral architectures.\\

\ACSection{CONCLUSIONS}
\noindent
In summary, we have developed the first systematic theoretical framework for understanding the CISS effect in dual-chiral supramolecular systems. Through quantum transport simulations of twisted multi‑layered helical assemblies, we demonstrate that the cooperative and competitive interplay between local molecular and global supramolecular chirality gives rise to several distinctive phenomena: enhanced spin polarization through chiral synergy, the coexistence of transverse and longitudinal spin filtering, and robust room-temperature performance facilitated by electron–phonon interactions.

The system exhibits marked field‑tunability. Under magnetic control, the interlayer twist breaks the structural symmetry and induces an anomalous angular phase shift in magnetoresistance, corresponding to a tilted spin‑filtering axis. This behavior establishes a vectorial optimization principle for spin injection, where efficiency is maximized by aligning the magnetization with this effective axis. Under optical excitation, the hierarchical chirality enables all‑optical spin polarization switching, breaking conventional enantiomeric symmetry. These findings not only provide the first theoretical foundation for CISS in multichiral architectures, but also establish hierarchical chirality as a versatile design principle for integrated molecular spintronics, enabling coupled control over optical, magnetic, and spin degrees of freedom.\\

\ACSection{ASSOCIATED CONTENT}
\noindent\SIicon\ \ACSSISubtitle{Supporting Information}\\
The Supporting Information is available free of charge at
https://pubs.acs.org/doi/***/***.

\vspace{1\baselineskip}
\ACSection{AUTHOR INFORMATION}
\noindent\hspace*{-0.3em}\ACSSISubtitle{Corresponding Author}
\begin{adjustwidth}{1.5em}{0pt} 
\textbf{Hua-Hua Fu} -- 
\textit{School of Physics and Wuhan National High Magnetic Field Center,
Huazhong University of Science and Technology, Wuhan 430074, People’s Republic of China;
Institute for Quantum Science and Engineering, Huazhong University of Science and Technology,
Wuhan 430074, People’s Republic of China;} %
\orcidicon~\href{https://orcid.org/0000-0003-3920-6324}{orcid.org/0000-0003-3920-6324};\,
Email: \href{mailto:hhfu@hust.edu.cn}{hhfu@hust.edu.cn}.
\end{adjustwidth}
\vspace{0.5\baselineskip}

\noindent\hspace*{-0.5em}\ACSSISubtitle{Notes}\\
The authors declare no competing financial interest.\\

\ACSection{ACKNOWLEDGEMENTS}
\noindent This work is supported by the National Key R$\&$D Program of China (2021YFC2202301).\\

\ACSection{REFERENCES}

\end{document}